\documentclass{iau}
\usepackage{graphicx}
\usepackage{natbib}

\def \arcmin     { ^{\prime} }

\def \hMpc      {h^{-1}{\rm\ Mpc}}

\newcommand{\persqdeg}{\ensuremath{\,\mathrm{deg}^{-2}}}
\newcommand{\lya}{Ly$\alpha$}
\newcommand{\dperp}{\ensuremath{\langle d_{\perp} \rangle}}
\newcommand{\sigthreed}{\ensuremath{\epsilon_{\rm 3D}}}
\newcommand{\ang}{\ensuremath{\mathrm{\AA}}}

\newcommand\apj{ApJ}%
\newcommand\aj{AJ}%
\newcommand\aap{A\&A}%
\newcommand\apjl{ApJL}%   
\newcommand\mnras{MNRAS}
\newcommand\apjs{ApJS}
\newcommand\jcap{JCAP}

\title[\lya\ Forest Tomography] %% give here short title %%
{\lya\ Forest Tomography of the \\ $z>2$ Cosmic Web}

\author[Khee-Gan Lee]   %% give here short author list %%
{Khee-Gan Lee
}

\affiliation{Max-Planck Institut f\"ur Astronomie, \\
K\"onigstuhl 17, Heidelberg, 69117, Germany}

\pubyear{2014}
\volume{308}  %% insert here IAU Symposium No.
\pagerange{119--126}
\jname{The Zeldovich Universe: Genesis \& Growth of the Cosmic Web}
\editors{A.C. Editor, B.D. Editor \& C.E. Editor, eds.}
\begin{document}

\maketitle

\begin{abstract}
The hydrogen \lya\ forest is an important probe 
of the $z>2$ Universe that is otherwise challenging to observe with galaxy redshift surveys, 
but this technique has traditionally been limited to 1D studies in front of bright quasars. 
However, by pushing to faint magnitudes ($g>23$) with 8-10m large telescopes it becomes possible to
exploit the high area density of high-redshift star-forming galaxies to create 3D tomographic maps of 
large-scale structure in the foreground. I describe the first pilot observations using this technique, 
as well discuss future surveys and the resulting science possibilities for galaxy evolution and cosmology.

\end{abstract}

\firstsection % if your document starts with a section,
\section{Introduction}

For decades, the hydrogen \lya\ forest absorption observed in the spectra of background quasars 
has been a crucial probe of the $z>2$ universe. 
Since the realization in the mid-1990s that residual neutral hydrogen in the photoionized intergalactic medium (IGM) is a non-linear tracer of 
the overall matter density distribution in the large-scale cosmic web \citep[the `fluctuating 
Gunn-Peterson' paradigm][]{cen:1994,bi:1995}, 
the \lya\ forest has enabled statistical studies of large-scale structure (LSS) at high redshifts. 

The \lya\ forest in each quasar spectrum is a one-dimensional probe of the foreground IGM,
 although a small number of close-pairs exist \citep[e.g.,][]{dodorico:2006}. 
 The area density of available quasars increase with limiting magnitude, and at faint magnitudes
 it becomes possible to study
 the 3D correlations of the \lya\ forest across different lines-of-sight: the BOSS \lya\ forest
 survey \citep{lee:2013} has observed $g \leq 21.5$ quasars with an average area density 
 of $15\,\mathrm{deg}^{-2}$, enabling measurement of the large-scale two-point \lya\ forest correlation function in 
 3D \citep{slosar:2011} and subsequently the detection of the $\approx 100\,\hMpc$ baryon acoustic oscillation scale 
 \citep{slosar:2013, delubac:2014}, which is useful as a cosmological distance measure.

 Beyond clustering studies, it is also possible to directly interpolate separate
 \lya\ forest sightlines to create a 3D map of the hydrogen absorption. This  
 `\lya\ forest tomography' was conceptually first proposed by \citet{pichon:2001} \citep[see also][]{caucci:2008},
 although the first 
 pilot observations were not conducted until recently \citep{lee:2014a}.
  In this article we will give a brief overview of this exciting new technique in terms of feasibility
  and initial pilot observations, as well as discuss future prospects.
 
 \section{Feasibility}
In simplest terms, \lya\ forest tomography simply interpolates between a set of \lya\ forest spectra to 
create a map of the foreground IGM absorption. Therefore the typical transverse separation, $\dperp$, of the
background sources must match the spatial scale of the desired map, \sigthreed. 
At $z=2.3$, the typical transverse separation of the BOSS \lya\ forest sightlines is $\dperp\sim 20\,\hMpc$ which is larger
than most scales of interest,
although there are efforts to create tomographic maps with BOSS \citep{cisewski:2014}.

The $\dperp$ is reduced by including fainter background sources, but the quasar luminosity function
is relatively shallow \citep{palanque-delabrouille:2013} and even at $g \leq 24$, the typical sightline separation
probed by quasars is $\dperp \sim 15\,\hMpc$.  However, at $g\gtrsim 23$, star-forming
galaxies (SFGs) begin to dominate the overall UV luminosity function \citep{reddy:2008},
allowing very dense grids to sample the foreground forest: at $g \leq [24,25]$ the sightline separations are $\dperp \approx [4,1]\,\hMpc$, respectively.

The question now turns to the requirements in 
terms of spectral resolution and signal-to-noise (S/N) in the data. If high-resolution 
($R\sim 30,000$, S/N$\gtrsim 20$ per pixel) spectra are needed, then 
\lya\ tomography will be a limited `novelty' technique even with 30m class telescopes, due to the extreme
exposure times ($>10$hr) needed to obtain such data with $g \sim 24$ sources.
Conversely, relatively noisy (S/N of a few) moderate-resolution spectra ($R \sim 200 - 1000$) 
are already achievable with existing 8-10m telescopes. 

In \citet{lee:2014}, we studied this issue using both simulations and analytic calculations. 
In terms of spectral resolution, the desired reconstruction scale, $\sigthreed$, must be resolved
along the line-of-sight (LOS), therefore one needs merely $R \geq 1300 (1\,\hMpc/\sigthreed)\allowbreak [(1+z)/3.25]^{-1/2}$. 
As for the necessary S/N, the requirements depend on $\sigthreed$ as well 
as the desired map fidelity. 
Since there will always be reconstruction errors on scales approaching the skewer separation, $\dperp$, 
due to the finite transverse sampling (`aliasing'), the individual spectra do not need
to oversample the absorption along the LOS. In simulated reconstructions using noisy mock data, we found that
$\mathrm{S/N} = 4$ per $\AA$ at a survey limit of $g=24.2$ is sufficient to give tomographic maps that look
similar to the `true' simulated field on scales of $\sigthreed \approx 3.5\,\hMpc$. 
This is achievable with existing telescopes: e.g. with the VLT-VIMOS 
spectrograph \citep{le-fevre:2003} such data would require 6hr exposure times per pointing.
The requirements do increase if we desire greater spatial resolutions: \lya\ forest tomography resolving
 $\sigthreed \lesssim 1\,\hMpc$ (i.e. $\lesssim 400$kpc physical at $z\sim 2.5$) would require 
$g >25$ background galaxies, which are accessible only with future 30m-class
telescopes.
 
 \section{Pilot Observations}
 In March 2014, we obtained moderate-resolution ($R \approx 1000$) spectra of 24 SFGs at redshifts $2.3 \lesssim z \lesssim 3$, 
 down to limiting magnitudes of $g\sim 24.7$, using the LRIS spectrograph on the Keck I telescope on Mauna Kea, Hawaii.
 These SFGs were all within a $5\arcmin \times 14\arcmin$ area of the COSMOS field \citep{scoville:2007}, resulting in an effective
 area density of $\sim 1000 \,\persqdeg$ or a nominal transverse separation of $\dperp \approx 2.3\,\hMpc$.
 
 The data was taken with $\approx 2$hr exposure times resulting in 
 $\mathrm{S/N} \sim 1.5 - 4$ per pixel. 
 It may seem aggressive to use such noisy data, 
 but our tomographic reconstruction uses a Wiener filtering algorithm \citep{wiener:1942} that incorporates noise-weighting. 
 Another concern is the possible contamination by intrinsic SFG absorption 
 lines, but studies of composite SFG spectra \citep{shapley:2003,berry:2012} have revealed a 
 relative lack of strong intrinsic absorption lines in the relevant $1040-1180\,\ang$ restframe region 
 once the foreground \lya\ forest fluctuations are averaged out by stacking.
 Moreover, UV spectroscopic observations of a small number of low-$z$ SFGs
 \citep[e.g.][]{heckman:2011}, in which the \lya\ forest is negligible, have revealed a relatively 
 flat `continuum', with only 2-3 strong absorbers that are straightforward to mask. 
 This is supported by the high-resolution spectrum of the
the lensed SFG cB158, which allowed Voigt-profile analysis of the
absorbers in its \lya\ forest region \citep{savaglio:2002}.

  \begin{figure}[t]
% \vspace*{-2.0 cm}
\begin{center}
 \includegraphics[width=\textwidth]{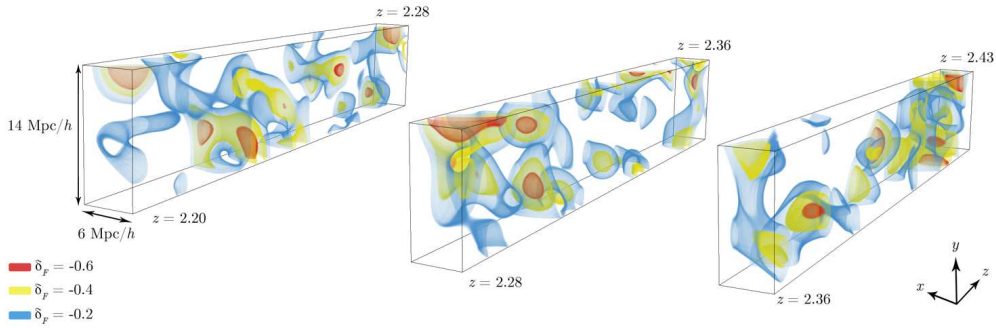} 
% \vspace*{-1.0 cm}
 \caption{3D visualization of the \lya\ forest tomographic map obtained from the foreground absorption 
 in 24 background SFG spectra, split into 3 different redshift segments for clarity. The color scale represents the
 relative \lya\ absorption, such that lower values correspond to higher densities. Excerpted from \citet{lee:2014a}.}
   \label{fig1}
\end{center}
\end{figure}

In Figure~1 we show the resulting 3D tomographic \lya\ forest reconstruction
\citep[see][]{lee:2014a}, spanning $2.2\leq z \leq 2.45$ with a spatial resolution of 
$\sigthreed \approx 3.5\,\hMpc$. The elongated map geometry is because bad weather
limited us to a small transverse area, whereas each sightline probes a long pathlength
$\sim 300\,\hMpc$ along the LOS.

Despite the narrow geometry, one sees large coherent structures spanning $\gtrsim 10\,\hMpc$ 
across the entire transverse dimension. These structures are contributed by multiple sightlines 
and are therefore
 unlikely to be due to random noise fluctuations nor
 intrinsic absorption (since the background SFG redshifts are not aligned). 
Simulated reconstructions on mock data with the exact same spatial sampling and S/N distribution yield a
good recovery of LSS features on scales of a few Mpc, although there are distortions on smaller scales.
We also compared the
map with a small number of coeval galaxies with known spectroscopic redshifts 
\citep[primarily from zCOSMOS,][]{lilly:2007}. In \citet{lee:2014a} we have shown that the galaxies live
preferentially in lower-flux regions (i.e. overdensities), although this correlation is weakened somewhat
by redshift errors on the coeval galaxies and tomographic reconstruction errors.
 
 \section{Science Applications \& Future Prospects}
 Our pilot observations have established the feasibility of \lya\ forest tomography,
 and motivates the COSMOS Lyman-Alpha Mapping and Tomography Observations (CLAMATO) survey, 
 which aims to obtain $\sim 1000$ SFG spectra within $\sim 1\mathrm{deg}^2$ of the COSMOS field, 
 aimed at creating a tomographic reconstruction of the \lya\ forest absorption at $\langle z \rangle \sim 2.3$
 over a comoving volume equivalent to $\sim (100\,\hMpc)^3$.
 
This survey will require $\sim 15$ nights of large-telescope time, and
will have various science applications: \textbf{(I)} The large-scale morphology and topology of the 
cosmic web has never been studied at $z \gtrsim 1$, and the tomographic map will allow us to study this 
on scales of $\gtrsim 3-4\,\hMpc$. \textbf{(II)} Using \lya\ forest absorption as a proxy for the
underlying density field, we will be able to study the properties of $z\sim 2$ coeval galaxies as a function of their
large-scale environment, yielding unique insights into this crucial era in galaxy formation and evolution.
\textbf{(III)} Within the volume covered by the map, we expect to find $\sim 10$ progenitors of massive 
$M>10^{14}M_\odot$ galaxy clusters. At $z \gtrsim 2$, these protoclusters are expected to manifest themselves
as overdensities of a few on spanning $\sim 10\,\hMpc$ \citep[c.f.][]{chiang:2013}, which should be easily 
detectable with \lya\ forest tomography. Finally, \textbf{(IV)} the data will allow us to probe 3D small-scale
\lya\ forest clustering on scales of $\lesssim 10\,\hMpc$, complementary to the larger scales probed by the BOSS
survey \citep{slosar:2011}. These measurements will be valuable in arriving at a comprehensive model of the \lya\ forest
absorption.
 
  \begin{figure}[t]
% \vspace*{-2.0 cm}
\begin{center}
 \includegraphics[width=\textwidth]{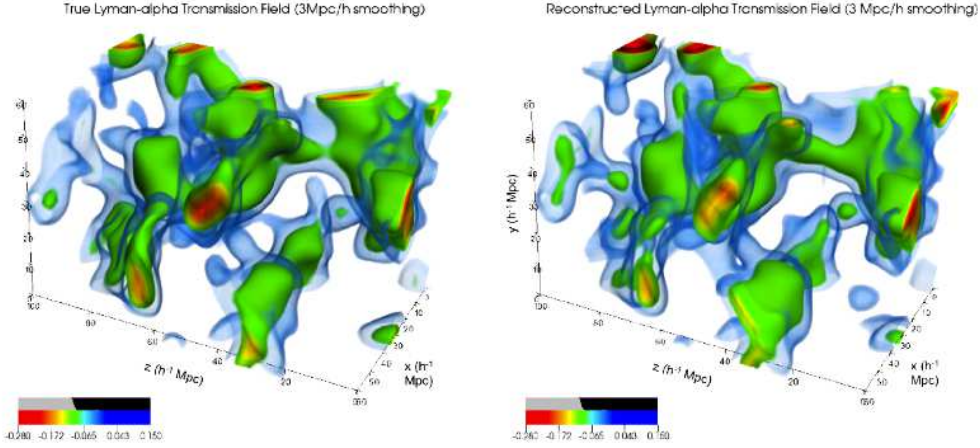} 
% \vspace*{-1.0 cm}
 \caption{(Right) Tomographic reconstruction using realistic mock data simulating the $1\,\mathrm{deg}^2$ CLAMATO
 survey, compared with the true 3D absorption field (left). Both fields have an effective smoothing of $\sigthreed = 3\,\hMpc$. The simulated survey data clearly recovers the LSS in the volume. Note that the $z$-dimension here
 is the LOS dimension, and in the real survey will span $\sim 2-3\times$ the distance shown here.}
   \label{fig2}
\end{center}
\end{figure}

%\bibliographystyle{plain}
%\bibliography{lyaf_kg,apj-jour,lss_galaxies}

\end{document}